# Ligand-dependent nano-mechanical properties of CdSe nanoplatelets: calibrating nanobalances for ligands affinity monitoring.

*Quentin Martinet[1], Justine Baronnier[1], Adrien Girard[2], Tristan Albaret[1], Lucien Saviot[3], Alain Mermet[1], Benjamin Abecassis[4], Jérémie Margueritat[1]\* and Benoît Mahler[1]\*.*

[1]Institut Lumière Matière Université de Lyon, Université Claude Bernard Lyon 1, UMR CNRS 5306, F-69622 Villeurbanne, France

[2]Sorbonne Université, CNRS UMR8233, MONARIS, Paris, France

[3]Laboratoire Interdisciplinaire Carnot de Bourgogne UMR 6303 CNRS-Université de Bourgogne Franche-Comté, 9 avenue A. Savary, BP 47870, 21078 Dijon Cedex, France

[4]Univ Lyon, ENS de Lyon, CNRS, Université Claude Bernard Lyon1, Laboratoire de Chimie UMR 5182, F-69342 Lyon, France

**Abstract:**

The influence of ligands on the low frequency vibration of different thicknesses cadmium selenide colloidal nanoplatelets is investigated using resonant low frequency Raman scattering. The strong vibration frequency shifts induced by ligand modifications as well as the sharp spectral linewidths make low frequency Raman scattering a tool of choice to follow ligand exchange as well as the nano-mechanical properties of the NPLs, as evidenced by a carboxylate to thiolate exchange study. Apart from their molecular weight, the nature of the ligands, such as the sulfur to metal bond of thiols, induces a modification of the NPLs as a whole, increasing the thickness by one monolayer. Moreover, as the weight of the ligands increases, the discrepancy between the massload model and the experimental measurements increase. These effects are all the more important when the number of layers is small and can only be explained by a modification of the longitudinal sound velocity. This modification takes its origin in a change of lattice structure of the NPLs, that reflects on its elastic properties. These nanobalances are finally used to characterize ligands affinity with the surface using binary thiols mixtures, illustrating the potential of low frequency Raman scattering to finely characterize nanocrystals surfaces.

**Introduction**

Vibrations in nanostructures are a powerful tool to study and characterize colloidal nanocrystals and heterostructures, especially quantum dots. The frequency of optical phonons is characteristic of the bonds in the material composing the nanocrystal, and is sensitive to their local environment.[1–9] Raman spectroscopy can therefore be an efficient tool to probe several features such as local defects, deformations induced by strain, as well as nanoscale alloying.[5,10] Furthermore, a careful analysis of Raman spectra of heterostructured nanocrystals can reveal in detail the nature of the heterostructure, the presence of sharp interfaces, interfacial alloying and electron-phonon coupling.[4,11] Phonons usually scrutinized in these cases are optical phonons, at frequencies higher than 100 $cm^{-1}$.

Lower frequency vibrations also exist and are characteristic of deformations of the nanocrystal as a whole. They are related to propagative acoustic phonons in the bulk material whose velocity, the sound velocity, is directly related to the physical properties of the material, namely its mass density and elastic constants. For small enough sizes, the vibrations are confined inside the nanocrystals and some of them can inelastically scatter light which manifest as peaks in the low-frequency Raman spectra. They are known as Lamb modes.[12] Their frequencies depend on the physical properties of the material as in the case of the acoustic waves, but also on the shape, size and environment of the nanocrystal. They can therefore provide crucial information about the shape of heterostructures in a non-destructive manner, as demonstrated in the case of dot-in-rods CdSe/CdS nanocrystals.[10]

Recent reports about low frequency phonons in nanocrystals[13–17] measured through diverse techniques (pump-probe, Raman) have also shown the importance of the surface energy of the nano-object. Indeed, controlling the surface states is crucial to control electron-phonon interactions occurring at nanocrystals surfaces. However, if the effect of the ligands on the electronic properties has been widely investigated, the mechanisms involved in the modifications of the phonon's frequencies are still unclear, especially in the case of the Lamb modes. This is due to two main reasons. First, the size of the nano-object must be small to observe a measurable effect of the ligands.[13,15,18] Second, studying small nanoparticles requires to study assemblies of nanoparticles presenting a small size and shape dispersion. Indeed, the width of the measured low frequency Raman peak is rapidly enlarged with these morphology parameters dispersion[19] and the effect of the ligands might be obscured. We have shown that nanoplatelets (NPLs) of CdS and CdSe are ideal objects to investigate this effect.[15,16] Their thicknesses are perfectly controlled at the atomic scale from 3 to 5 atomic layers (0.9 nm to 1.5 nm in case of CdSe), and due to their lateral dimensions which is much larger than their thickness, only the breathing vibration of the thickness is measured, making them model nano-objects with negligible size dispersion. Moreover, in these 2D systems, the ligands nature can be easily modified through standard ligands exchange procedures.[20–22]

We demonstrated previously[15] that in the framework of continuum elasticity, the frequency of the breathing vibration of a free CdSe NPL depends on its density $\rho$ and elastic constant $C$ ($C_{11}$ in the case of cubic structure) according to

$$\nu = \frac{1}{2h}\sqrt{\frac{C}{\rho}} \quad (1)$$

where $h = (a_0 \times \text{number of atomic layer})$ is the thickness of the NPL expressed as a function of the lattice parameter $a_0$ and $\sqrt{\frac{C}{\rho}} = v_L$, i.e. the longitudinal speed of sound in the NPL. The presence of the native ligands can be taken into account through the inertial mass load they induce.[15,16] The lumped mass effect is accounted for with the following equation, whose roots give the resonance frequencies of the NPL across the thickness:

$$\cos\left(\frac{\omega}{V_L}\frac{h}{2}\right) = \frac{\sigma\omega}{\rho V_L}\sin\left(\frac{\omega}{V_L}\frac{h}{2}\right) \quad (2)$$

where σ is the surface mass density of the ligands and ρ the volume mass density of the NPL. The magnitude of this effect is driven by the ratio of the masses between the ligands and the NPL (σ/ρ factor).

In this article, we demonstrate that other parameters play a role in the change of frequency and that an improved model is required. We show that the breathing vibration of NPLs also depends on the structural modification of the NPLs induced by the presence of ligands at their surfaces. This provides a new mean to control phonons in such nanostructures but also a new tool to characterize ligand induced modification of their lattice structure. This latter effect is all the more important when the NPLs are thin, i.e. when the proportion of atoms close to the surface is large. Significant deviations from the mass-load model are reported. They are explained by the modification of the NPL structure as a function of ligand nature, which impacts on the mechanical properties of the NPLs. The fine characterization of this modification allows to calibrate the NPLs vibration frequency as a function of the molecular weight on their surface using the experimental measurements, and the ability of these calibrated nanobalances is demonstrated by studying the relative binding affinities of different thiol ligands at the NPL surface.

**Results**

   **A. Nanobalance calibration**

In order to characterize the sensitivity of the breathing vibration of the NPLs to the mass of the ligands present on its surface, we prepared CdSe NPLs with a 5 monolayers thickness. After synthesis, the surface of the NPLs is covered with oleic acid ligands (OA, molecular weight: 282.46 g.mol$^{-1}$) and the black curves on figures 1a and 1b show the corresponding UV-Visible absorption and low frequency Raman spectra respectively. The initial absorption spectrum shows two peaks at 515 and 550 nm that are the typical signature of the exciton confinement in a 5 monolayer NPL with oleic acid ligands.[23] After ligand exchange by alkyl-thiols (see material and methods), the excitonic energy is reduced, which results in a red shift of the excitons peaks to 536 and 567 nm respectively, confirming that the ligand exchange is complete.[20,24] The length of the thiol chain, varied to modify the mass of the ligands from 146.29 g.mol$^{-1}$ for octanethiol (OT), to 286.56 g.mol$^{-1}$ for octadecanethiol (ODT), does not influence significantly the energy of the exciton. To measure the vibration frequency as a function of the ligand molecular weight, the measurements must be performed in close resonance with the lower energy exciton peak.[15] In such condition a strong luminescence background overwhelming the Raman signal is observed, and a low concentration of copper ions are introduced in the NPLs in order to quench part of the luminescence (see methods section).

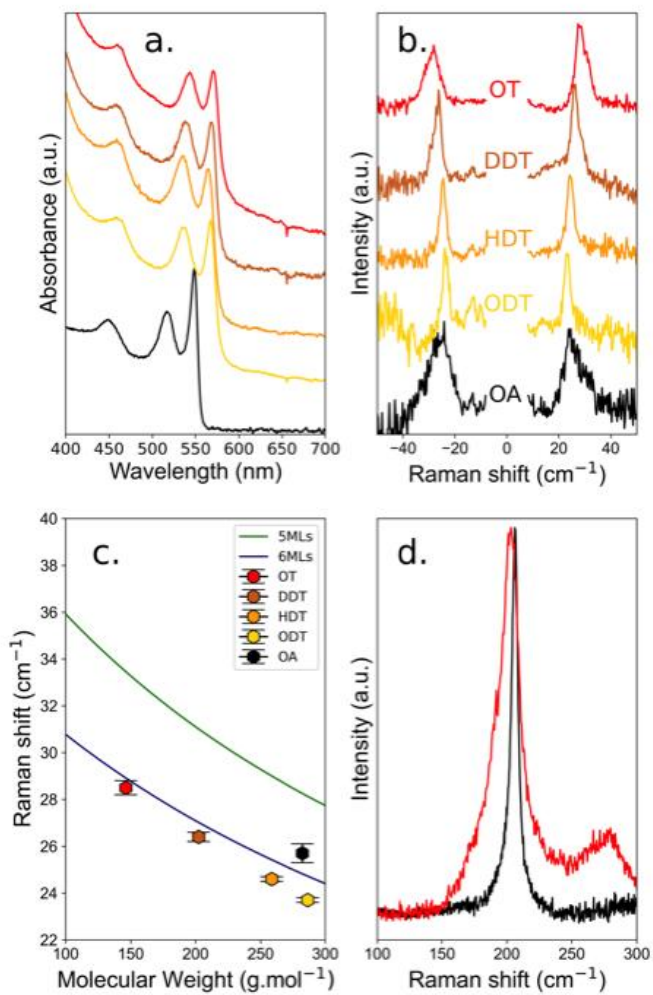

**Figure 1: a.** Optical absorption spectra of 5ML NPLs as a function of the ligand type (from top to bottom: OT, DDT, HDT, ODT and OA). **b.** Corresponding low frequency Raman spectra (the central line corresponding to the Rayleigh scattering has been removed for clarity) **c.** Comparison of the experimental frequency with the mass load model.[15] **d.** Raman spectra of NPLs with OA ligands (black) and NPLs with OT ligands (red).

Figure 1b shows the variation of the breathing vibration frequency as a function of the thiol ligand molecular weight. An increase of the ligand molecular weight induces a down-shift of the frequency (from 28.7 cm$^{-1}$ to 23.5 cm$^{-1}$), as predicted in our previous work.[15] However, the vibration frequencies measured for OA and ODT whose masses are almost the same (see methods section), are significantly different: 25.7 cm$^{-1}$ and 23.5 cm$^{-1}$ respectively. This behavior is reported in figure 1c, where the experimental results (hexagonal dots) are compared with the mass-load model[15] for 5 monolayers and 6 monolayers thick NPLs. Surprisingly, the measured frequency shifts as a function of thiol molecular weights are in good agreement with the model only if we consider 6ML NPLs while they are in between 5 and 6 monolayers for OA ligands. The influence of thiol ligands at the surface of CdSe nanocrystals has already been investigated. This type of ligands induces a red-shift of the lowest excitonic transition of CdSe nanocrystals,[25–27] as shown in Figure 1a for our NPLs. This shift has first been interpreted as a decrease of the quantum confinement of the hole due to the presence of the ligands.[25–27] But this effect can also be explained if one considers that the sulfur atoms of the thiol ligands are forming a shell around the CdSe nanoplatelets thus increasing slightly its effective thickness. Then it is the delocalization of the exciton that induces a red-shift of the excitonic peak.[5] In the case of NPLs, considering that the sulfur atoms coming from the thiol ligands are part of the lattice structure, half a monolayer is added on each side of the NPLs, increasing the total thickness by one monolayer. If we consider this hypothesis with the mass load model, the agreement with the experimental measurements is significantly improved for thiol molecules as shown in Figure 1c. The presence of this sulfide layer is supported by the Raman spectra at higher frequency (between 200 and 300 cm$^{-1}$) where the longitudinal optical phonons of CdS and CdSe manifest. Figure 1.d shows the Raman spectrum measured on the sample of NPLs covered with OT ligands. In this spectrum, an intense asymmetric peak at 203 cm$^{-1}$, and a band around 275 cm$^{-1}$ are observed. The first peak corresponds to the CdSe LO phonon. The broadening of the peak with respect to bulk CdSe LO phonon is explained by the increasing role of surface optical phonons (at lower frequency) and by the spatial confinement that induces scattering from phonons (higher frequency).[28–30] The second band has already been observed[5,24,29–32] when performing resonant Raman measurement on CdSe QDs plus CdS or elemental S. It was assigned to the presence of sulfur atoms that form CdS bonds whose LO phonon frequency depends on the thickness of the CdS layer. For a single monolayer the frequency of this CdS-like LO phonon is around 270 cm$^{-1}$. The difference in frequency between the CdSe LO phonon and the CdS-like LO phonon can be used to estimate the quantity of Se atoms with respect S atoms.[33] Here a difference of 72 cm$^{-1}$ is measured which approximately corresponds to 80% of Se and 20% of S atoms. This is the expected ratio if we consider NPLs of 5MLs with half monolayer of sulfur on each sides. Unfortunately, this band was not observed on all samples due to the luminescence background but the effect of the sulfur bonds is also clearly evidenced at low frequency when considering the breathing vibration mode of the NPLs.

Indeed, the origin of this vibration mode at low frequency has also been discussed by A. I. Lebedev in ref. 34. In this publication the author uses first-principle calculations and modeling of the Raman spectra to determine the origin of this vibrational peak. CdSe NPLs of different thicknesses (2 to 6 MLs) are modeled with a fluorine atoms layer compensating the cadmium atoms surface charge for the calculations. Vibration modes with a displacement out of plane, qualified as quasi-Lamb, are calculated but their frequencies are not in good agreement with our experimental frequencies for NPLs covered with OA. However, it is interesting to note that fluorine atoms are placed in the vacant selenium positions, and that the wave confinement calculated in 34 is defined on the thickness of the plate plus the two layers of fluorine atoms. These theoretical frequencies are therefore obtained for a bare platelet of CdSe with an additional layer of fluorine atoms and must be compared to the Lamb model of a bare NPL with the same thickness. When considering this additional monolayer, a good agreement between the approach in ref 34 and the Lamb model is observed. These results thus confirm that the sulfur atom constituting the thiol group is involved in a molecular bond with the cadmium atoms present at the surface of the NPLs.

This influence of the binding group explains why the NPLs with OA ligands do not follow the same behavior as the NPLs with thiol ligands. However, in the case of OA, there is still a discrepancy between the mass-load model for 5 monolayers and the experimental measurement. Due to the preparation method, the NPLs surface is almost completely covered with carboxylates[35], that is a mixture of oleate and acetate ligands (in unknow proportions). The overall molecular weight to be taken into account should then be lower and may lead to a better agreement with the mass-load model. Moreover, one could expect that the oxygen atoms of the carboxylic function of OA might also be involved in the structure of the NPLs. To understand this behavior, thinner NPLs with thicknesses of 3 and 4 monolayers were synthesized with OA native ligands, and we carried out the same study with thiol ligand exchange.

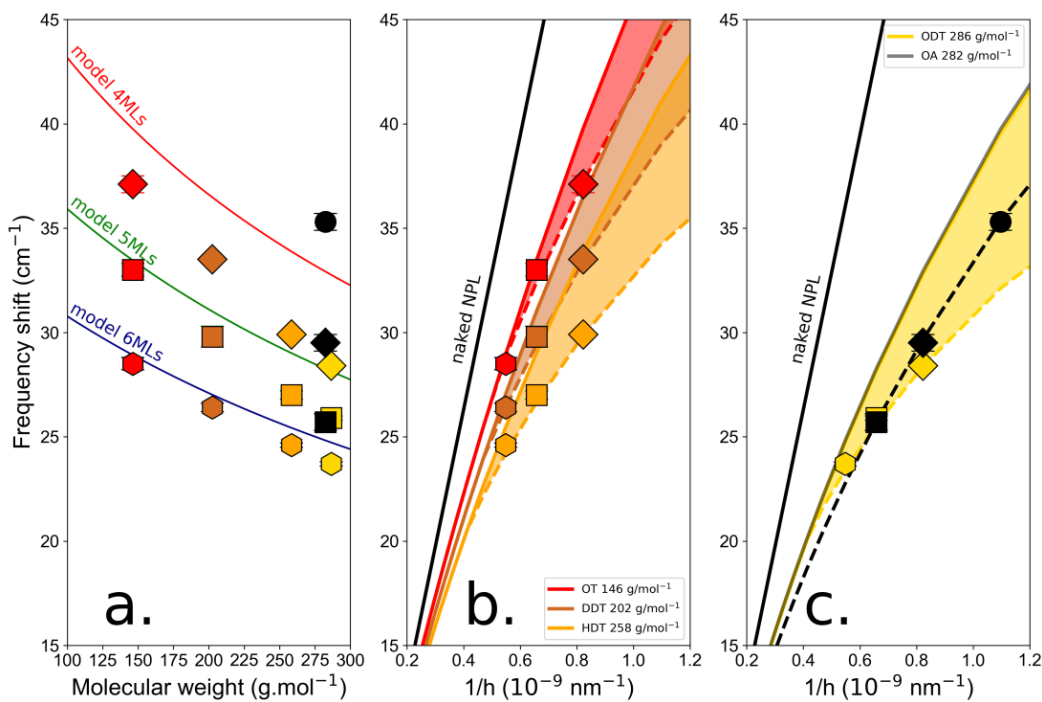

**Figure 2: a.** Measured frequencies of NPLs vibration modes as a function of thiol molecular weight for different NPL thicknesses. Vibration frequencies of the initial oleate NPLs are also included. Red, brown, orange, yellow and black symbols correspond to OT, DDT, HDT, ODT, and OA respectively. Hexagone, square, diamond, and circle symbols correspond to 6, 5, 4, and 3 MLs thick. The experimental data are compared with the mass load model[15] for the different thicknesses 6MLs (blue line), 5MLs (green line), 4MLs (red line). **b. and c.** Data for OT(red), DDT(brown), HDT(orange), ODT(yellow) and OA (black) are represented as function of the inverse thickness of the NPLs and compared to the theoretical value for a naked NPL (black curve) (figure b and c are splitted for clarity). Continuous lines are calculated using the mass-load model. The dotted lines are guide to the eye, that are obtained by considering a modification of the sound longitudinal sound velocity that depends on the molecular weight.

The measured frequencies of the breathing mode are reported in Figure 2 and compared with the mass-load model. The data are represented as a function of the molecular weight of the ligands (Figure 2.a) and as a function of the number of layers for fixed molecular weight (Figure 2b. and c.) As demonstrated previously, with thiols we have to consider that the thickness increases by one monolayer and thus if we compare NPLs with the same thicknesses we observe that the frequency shifts measured for OA and ODT are almost the same. This confirms that the sulfur atoms are part of the lattice structure of the NPLs.

It is clear from figure 2 that the increase of the thiol length, i.e. the surface mass experienced by the NPLs, induces, as expected, a decrease of the breathing vibration mode frequency for all the thicknesses considered. Nevertheless, the massload model only reproduces correctly the behavior observed for 6 MLs NPLs. When considering 5 MLs NPLs, the agreement is still good but a deviation from the model is observed for the heaviest molecules. When reducing the thickness to 4 MLs, the discrepancy between our model and the experimental data increases again. Figure 2b. and 2.c represent the same set of data as a function of the inverse thickness of the NPLs, and are compared with the ideal case of a bare NPL. For the thiol molecules, we clearly observe that the measured frequencies are lower than the frequency expected with the mass-load model. The difference between these frequencies increases when the molecular weight of the thiols increases, and this effect is all the greater as the NPL thickness is reduced. Figure 2.c also shows a comparison between ODT and OA molecules that have almost the same molecular weight. Albeit the general behavior is the same (experimental frequencies lower than expected frequencies and function of the NPL inverse thickness), the experimental frequencies for OA are closer to the mass-load model and follows a different behavior than ODT. Those observations suggest that the mass-load model is still incomplete for thin NPLs.

This model is based on several hypotheses: the first one is the surface coverage. To calculate the surface mass density of the ligands at the surface of the NPLs, we considered that each surface cadmium atom is linked to one ligand (surface density: 5.4 ligands.nm$^{-2}$). This is an overestimation because the typical values deduced by NMR are between 5.3 and 4.6 ligands.nm$^{-2}$ for OA ligands.[35,36] The effect of a lower surface covering results in an increase of the frequency shifts (Figure S1 of Supplementary Materials) which is not the experimental behavior. One would expect a different surface covering for thiol ligands, but thiol molecules have a greater affinity with nanoplatelets than OA, and therefore one can expect at least the same surface covering. Finally, while the surface covering of CdSe nanoplatelets with thiol molecules has not been measured, the surface covering of thiol molecules on gold surfaces has been extensively studied showing that the alkane chain length does not affect the covering.[37–39] These works also highlights the fact that the geometric effect can induce an increase in surface covering. In what follows, a fixed surface density of 5.4 ligands.nm$^{-2}$ is therefore considered whatever the nature of the ligand.

The other parameter that can be modified is the value of the sound velocity of the CdSe nanoplatelet, that is related to its elastic and structural properties. In our model we used the bulk value of CdSe considering the zinc-blende structure of the NPLs: the lattice parameter a = 0.608 nm and the theoretical estimated elastic constant $C = C_{11} = 88$ GPa, giving a longitudinal sound velocity of 3945 m.s$^{-1}$.[15,16] These value were deduced from the bulk material considering a zinc-blende structure, but it is an estimation. In particular, several works have shown that the NPLs crystalline structure is slightly different from zinc-blende structure.[21,40]

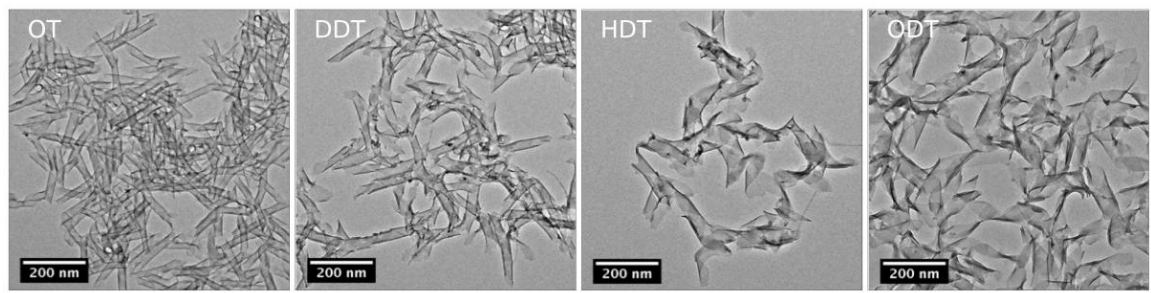

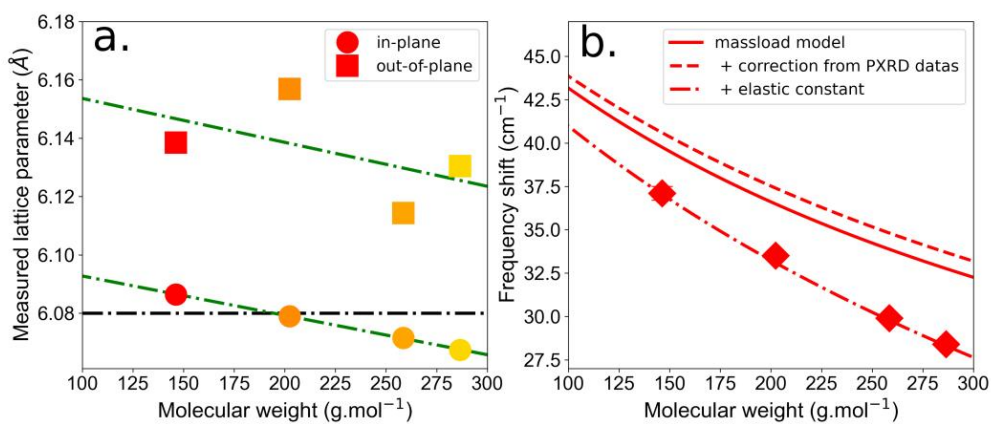

**Figure 3: Top.** TEM images of the NPLs (3+1MLs) morphology as a function of the ligand. **a.** In-plane and out-of-plane lattice parameters measured by XRD for the 4 monolayers NPLs as a function of the molecular weight of the ligands. **b.** The continuous red line corresponds to the mass load model considering a zinc-blende structure with a constant C=88 GPa and the bulk lattice parameter a = 6.08 angstroms. The doted red line considers the modification of the lattice parameter as a function of the ligand mass deduced from PXRD measurements, and the induced modification of the density $\rho$. Finally, the dashed red line considers an additional modification of the elastic constant C depending on the ligand molecular weight. This latter curve is used as a calibration curve of the nanobalances with 4MLs thickness.

The TEM images (figure 3) of the NPLs clearly show that ligands strongly impact their morphology. They induce a winding of the NPLs whose amplitude varies with the thiol ligand length. This behavior is unexpected and might be related to steric hindrance. Moreover, the nature of the bond between the molecule and the surface is also probably important to control the winding, but further experiments with different ligands of well-known steric hindrance and functional type should be performed to answer more specifically this question. Be it as it may, the TEM images shows a clear dependency of the winding radius as a function of the thiol length.

We performed PXRD measurements on our samples to determine the in-plane and out-of-plane lattice parameters (diffractograms in Supporting Information). Figure 3a confirms that the crystalline structure of the NPLs is more accurately described as pseudotetragonal, due to the differences between in-plane and out-of-plane lattice parameters.[40] The measured in-plane lattice parameter is slightly smaller (1%) than the out-of-plane parameter. Moreover, the in-plane and out-of-plane lattice parameters decrease slightly as a function of the alkyl chain length. It has to be noticed that due to the nano 2D nature of the sample, the peaks observed on the PXRD diffractograms and corresponding to the out-of-plane lattice parameter are difficult to fit and more precise PXRD measurements may be needed to refine the accuracy of the in plane and out of plane lattice parameters. Yet Figure 3a thus confirms that a modification of the lattice parameters and thus of the unit cell depends on the thiol chain length. This structure modification, from cubic to tetragonal, implies to take $C_{33}$ as elastic parameter along the c-axis of the NPLs (i.e. along the thickness), instead of $C_{11}$. Unfortunately, the elasticity of such structure at the nanoscale is poorly characterized and no value of $C_{33}$ is reported, a fortiori as a function of ligand mass. Therefore, we rely on an empirical dependence of $C_{33}$ as a function of ligand mass to calibrate our nanobalances. Thus, by fitting the experimental data with our mass load model considering the modification of the lattice parameter (and density $\rho$) (see Supporting Information) we can deduce, for the 4MLs NPLs, the dependency of the elastic parameter as a function of the ligand mass $C_{33} = C - 0.091 \times M$ ; where $C = C_{11,ZB} = 88$ GPa and M is the molecular weight in g/mol. This dependency is used in figure 3b to calibrate the behavior of our nanobalances. Although the linear dependency of the elastic parameter as a function of the unit cell deformation (and thus on the molecular weight) is expected, $C_{33,ZB}$ should be replaced by the value of the elastic constant of a bare NPLs of 4MLs. Unfortunately, this value is not yet available. In addition, this dependence is extracted from the experimental data considering a surface covering of 100%. This is a strong assumption which also means that the potential structural deformation of NPLs is potentially greater than we expected. Anyway, the nanobalances are simply calibrated using the experimental data and the unique unknow quantity is the surface covering which is considered constant regardless of the nature of the ligands to determine the difference in binding affinity between OT and ODT.

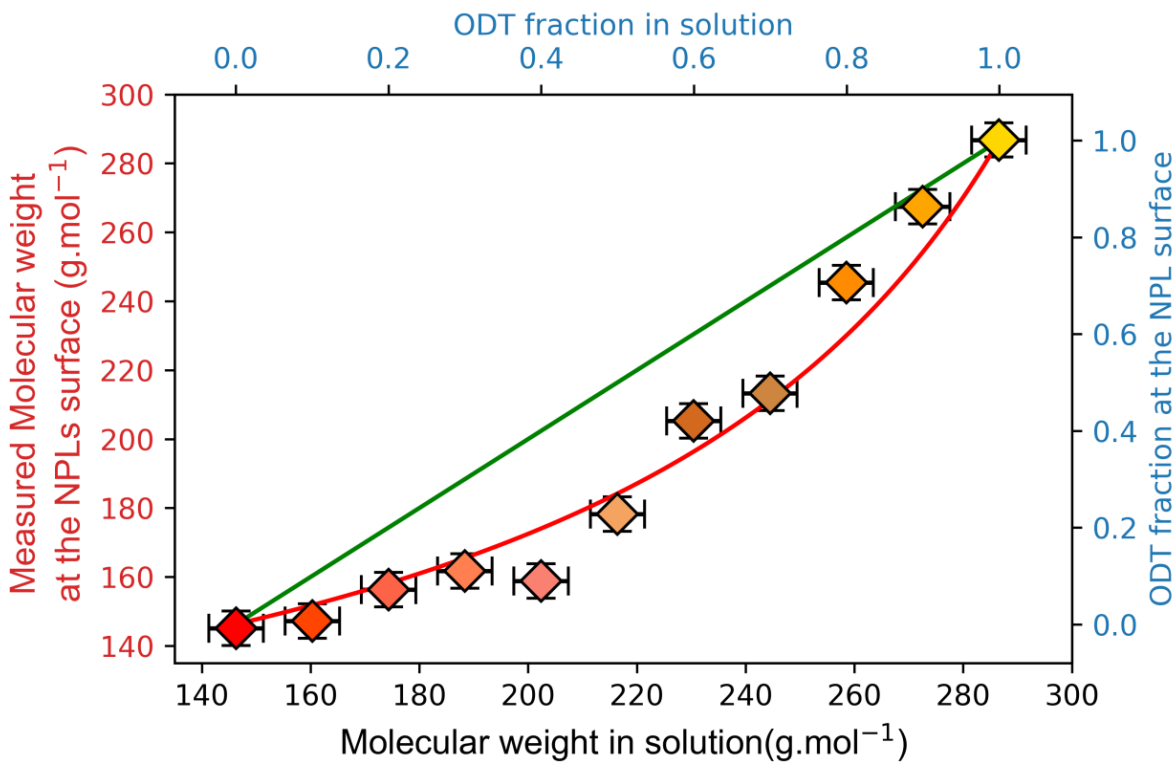

**Figure 4:** Competitive binding between octanethiol and octadecanethiol on 3ML CdSe NPLs using frequency shift measurement to determine the weight of the surface-bound ligands. The green and red curves are the result of the equilibrium model considering K = 1 and K = 2.7 respectively.

### B. Using the nanobalance effect to monitor the binding of ligands

In the previous part of the article, the study of the frequency shifts gave us insight about mechanical property variations of the nanoplatelets grafted with ligands of different chain length and binding groups that must be considered to calibrate the nanobalances. It has been possible to obtain an experimental relationship linking the frequency vibration of the NPL to the molecular weight of the ligands attached to the surface in the case of linear alkanethiols. We will now explore the possibility to use the NPL as a nanobalance that can allow to precisely analyze the composition of surface ligands in more complex cases where other techniques are not sensitive.

As we establish this relationship for linear alkanethiols we will consider the following equilibrium of binding between octanethiol and octadecanethiol at the surface of the nanoplatelets:

$Cd_{NPL}$- octadecanethiolate + OT $\rightleftharpoons$ $Cd_{NPL}$-octanethiolate + ODT

The equilibrium constant K is simply expressed by:

$$K_{\text{NPL-Thiols}} = \frac{[ODT][Cd_{NPL}OT]}{[OT][Cd_{NPL}ODT]}$$

We are expecting K to be near a value of 1, as the differences between the two considered ligands only reside in their chain length that shouldn't greatly modify their affinity towards the NPL cadmium surface cations. The experiment simply consists in mixing 3ML NPLs bearing their native carboxylate ligands with a large excess of an ODT:OT binary mixture of varying ratios. The measured spectra are shown as supplementary materials (figure S4) and figure 4 show the measured frequencies as a function of the molecular weight and ODT fraction. The ligand exchange is let to proceed at 65°C for 72h as described in the experimental section. Considering that the thiols are in large excess compared to the native carboxylates and that their affinity for the NPL surface is much higher, the observed equilibrium is practically the one noted above. If we call x the fraction of ODT in solution and y the fraction of ODT at the surface of the NPLs, the equilibrium can be written as:

$$K_{\text{NPL-Thiols}} = \frac{[ODT][Cd_{NPL}OT]}{[OT][Cd_{NPL}ODT]} = \frac{x}{(1-x)} \times \frac{(1-y)}{y}$$

Considering that the excess of free ligands is very large, y is then determined using the low frequency Raman measurement followed by the calculation of the average molar mass of ligands needed to reach this frequency according to the calibration curve (figure 3b). The results presented in figure 4 show a clear deviation to the linear case (K=1 green line), and the experimental values can be fitted reasonably well using our simple equilibrium model with a K = 2.7. This simple competitive binding experiment unambiguously demonstrate that OT has a higher affinity than ODT for the NPL surface cadmium ions and allow us to quantify the affinity difference. This result also allows to determine the energy difference between Cd-OT and Cd-ODT molecular bonds at the NPL surface. If we consider that:

$$K_{d(OT)} = C \times e^{(\Delta G_{OT-NPL}/RT)}$$

It is possible to determine the energy difference $\Delta G_{OT-NPL} - \Delta G_{ODT-NPL}$ since:

$$K_{\text{NPL-Thiols}} = \frac{[ODT][Cd_{NPL}OT]}{[OT][Cd_{NPL}ODT]} = \frac{K_{d(OT)}}{K_{d(ODT)}} = \frac{e^{(\Delta G_{OT-NPL}/RT)}}{e^{(\Delta G_{ODT-NPL}/RT)}} = e^{\frac{1}{RT}(\Delta G_{OT-NPL}-\Delta G_{ODT-NPL})}$$

We measured $K_{\text{NPL-Thiols}} = 2.7$ which gives an energy differences of $\Delta G_{OT-NPL} - \Delta G_{ODT-NPL} \approx RT = 2.48$ kJ/mol.

These results clearly show that the vibrations of nanoplatelets are a really sensitive signature of their structure as well as their surface ligands. Being able to disentangle these effects give access to a sensitive way to probe their mechanical properties and to follow their surface modifications.

In conclusion, we have here demonstrated the sensitivity of the breathing vibration mode of CdSe NPLs to ligands molecular weight. By carefully considering the real thickness of the NPLs, we highlight that for relatively thick (6MLs) NPLs this breathing mode depends on the ligands molecular weight and follows the behavior predicted by the mass-load model we previously developed. However, when reducing the number of layers to potentially increase the breathing mode sensitivity to ligands weight variations, we observed that other effects must be considered. At this scale, the structure of the whole NPL is impacted by the surface modifications induced by the ligands. This results in a modification of the NPL mechanical properties that strongly depend on the nature of the ligands. To understand the dependency of the mechanical properties of the NPL plus ligands system further experimental and theoretical investigations must be performed. Nevertheless, we were able to calibrate the nanobalances based on the relationship between the NPL vibration frequency and the molecular weight of ligands of the same type (alkyl chains), which was deduced experimentally. This calibration has finally been applied to monitor the competitive binding at the surface of 3ML NPLs between thiols of different chain length. Although the development of more comprehensive models is still needed, the ability to obtain information on the structural properties by simply using inelastic light scattering spectroscopy opens new routes to investigate the physical-chemistry of the interfaces, and the effect of surface treatments on nano-objects. Finally, these results open a route to develop a new class of label-free sensors based on a nanoplatelet acting as a nanobalance probed through inelastic light scattering.

**Methods**

Chemicals:

1-octadecene (ODE, 90%), oleic acid (90%) and 1-octylamine (99%) were purchased from Alfa Aesar. 1-octanethiol (OT, 98.5%), 1-octadecanethiol (ODT, 98%), 1-dodecanethiol (DDT, 98%), 1-hexadecanethiol (HDT, 95%), trioctylphosphine (TOP, 97%), tetrakis(acetonitrile)copper(I) hexafluorophosphate (97%), cadmium nitrate tetrahydrate (Cd(NO$_3$)·4H$_2$O, 99%) and cadmium acetate dihydrate (98%) were obtained from Sigma-Aldrich. Selenium powder (99.99%) was purchased from Strem. Na(myristate) (98%) was purchased from TCI.

All chemicals were used directly without purification.

Precursors:

Cd(myristate)2:

9.3 g of Na(myristate) is dissolved in methanol under vigorous magnetic stirring for 1 h. 3.7 g of Cd(NO$_3$)·4H$_2$O dissolved in 120 mL MeOH is slowly added to this solution inducing Cd(myr)$_2$ precipitation. After one more hour stirring, the white solid is filtered and washed multiple times on Buchner.

3MLs CdSe NPLs synthesis:

14 mL ODE, 185 mg Cd(Acetate)$_2$·2H$_2$O and 190 μL oleic acid are introduced in a 100 mL three neck flask. The temperature of the solution is raised to 220°C under Argon flux. After injection of 300 μL of TOPSe 1M, the reaction mixture is annealed at the same temperature for 10 minutes resulting in the formation of 3 MLs CdSe NPLs. At the end of the reaction, 2 mL of oleic acid is added before the solution is let to cool down to room temperature. NPLs are isolated from the reaction mixture by centrifugation; they are then dispersed in hexane and precipitated one more time by addition of ethanol. The final product is dispersed in 10 mL hexane for storage.

4MLs CdSe NPLs synthesis:

85 mL ODE and 1.02g Cd(myr)$_2$ are introduced in a 250 mL three-neck flask and degassed 1h under vacuum at 100°C. Under Argon, 72 mg Se powder dispersed in 5 mL ODE is introduced in the flask. The temperature of the setup is set to 240°C and 300 mg Cd(Acetate)$_2$ is introduced in the flask at 205°C. The reaction mixture is annealed at 240°C for 12 minutes resulting in the formation of 4 MLs CdSe NPLs. At the end of the reaction, 6 mL oleic acid is added before the solution is let to cool down to room temperature. NPLs are isolated from the reaction mixture by precipitation using a hexane/ethanol mixture and centrifugation; they are then dispersed in hexane and precipitated one more time by addition of ethanol. The final product is dispersed in hexane for storage, adjusting the solution volume to obtain an optical density of 1.5 at 350 nm when diluted 15 times.

5MLs CdSe NPLs synthesis:

56 mL ODE and 680 mg Cd(myr)$_2$ are introduced in a 250 mL three-neck flask and degassed 1h under vacuum at 100°C. The temperature of the solution is raised to 250°C under Argon flux and 48 mg of Se powder dispersed in 2ML ODE is swiftly injected at this temperature. Immediately after injection, 480 mg Cd(Acetate)$_2$ is introduced in the flask. The reaction mixture is annealed at the same temperature for 15 minutes resulting in the formation of 5 MLs CdSe NPLs. At the end of the reaction, 10 mL oleic acid is added before the solution is let to cool down to room temperature. NPLs are isolated from the reaction mixture by centrifugation; they are then dispersed in hexane and precipitated one more time by addition of ethanol. The final product is dispersed in hexane for storage, adjusting the solution volume to obtain an optical density of 1.5 at 350 nm when diluted 15 times.

Ligand exchange:

The ligand exchange protocol is adapted from ref([20]). Typically, 200 μL of the as-synthetized NPLs in hexane are mixed with 2 mL hexane and 1 mmol alkanethiol (octanethiol, dodecanethiol, hexadecanethiol or octadecanethiol). The mixture is then placed in an oven at 65 °C for 72 h to allow for a complete ligand exchange to occur. After exchange, the NPLs are washed trice by precipitation with ethanol, centrifugation and dispersion by sonication in 5 mL hexane.

Copper doping:

A stock solution of tetrakis(acetonitrile)copper(I) hexafluorophosphate, $[MeCN]_4Cu^IPF_6$, at $1 \times 10^{-4}$ M in ethanol is used to dope the NPLs. The stock solution of copper (I) ions is directly added to NPLs in hexane at a ratio of 1 Cu for 1000 Cd, and the mixture is heated at 65°C for 1h. Excess ions are then discarded by precipitation in ethanol, centrifugation and dispersion in hexane. This copper doping allows to quench part of the luminescence in order to observe the Raman scattering signal.

Absorption spectra: Absorbance spectra have been recorded using an AvaSpec-ULS2048 fiber optic spectrometer equipped with an AvaLight balanced Deuterium-Halogen lightsource.

PXRD: Powder X-ray diffraction (PXRD) patterns were recorded on a PANalytical Empyrean diffractometer (Cu Kα radiation). Samples were prepared by drying on a low background silicon substrate.

Raman spectroscopy: Resonant Raman spectra were obtained using Labram HR and Renishaw microspectrometers equipped with ultra-low frequency notch filters. The spectra were acquired using different laser lines (532, 633, and 660nm) depending on the thicknesses of the NPls and thus on the energy of the excitons. A long working distance x50 objective was used and each time the laser power was fixed to avoid any damage on the sample for several hours (typical: 10 mW). Several spectra were acquired on different positions of the same samples and were used to extract the frequencies using a gaussian fitting method.

**ACKNOWLEDGMENT**

This work was partially supported by the ANR NanoVip project, Grant ANR.13.JS10.0002 of the French Agence National de la Recherche, the Fédération André Marie Ampère 2013 (FRAMA) and the Programme Avenir Lyon Saint-Etienne from Université de Lyon in the framework "Investissements d'Avenir" (ANR-11-IDEX-0007). L.S. acknowledges support by EUR-EIPHI Graduate School (contract ANR-17-EURE-0002). The authors declare no competing financial interest. We thank the Consortium Lyon St-Etienne de Microscopie (CLYM) for access to the microscope platform TEM characterizations, and the spectroscopy platform CECOMO at Villeurbanne for the access to the LABRAM-HR spectrometer.